\documentclass{sig-alternate-05-2015}
\usepackage{todonotes}

\begin{document}

\setcopyright{acmcopyright}

\setcopyright{acmcopyright}

%
\conferenceinfo{\\Medical Information Retrieval Workshop (MedIR)}{July 21, 2016, Pisa, Italy}

\title{Implicit Negative Feedback in Clinical Information Retrieval}
%
%
%
%
%

\numberofauthors{2} 
%
\author{
\alignauthor
Lorenz Kuhn\\
      \affaddr{Department of Computer Science ETH Zurich}\\
      \affaddr{ETH Zurich, Switzerland}\\
      \email{kuhnl@student.ethz.ch}
\alignauthor
Carsten Eickhoff\\
      \affaddr{Department of Computer Science ETH Zurich}\\
      \affaddr{ETH Zurich, Switzerland}\\
      \email{ecarsten@inf.ethz.ch}
}

\maketitle
\begin{abstract}
In this paper, we reflect on ways to improve the quality of bio-medical information retrieval by drawing implicit negative feedback from negated information in noisy natural language search queries. We begin by studying the extent to which negations occur in clinical texts and quantify their detrimental effect on retrieval performance. Subsequently, we present a number of query reformulation and ranking approaches that remedy these shortcomings by resolving natural language negations. Our experimental results are based on data collected in the course of the TREC Clinical Decision Support Track and show consistent improvements compared to state-of-the-art methods. Using our novel algorithms, we are able to reduce the negative impact of negations on early precision by up to 65\%.
\end{abstract}

%
%

\ccsdesc[500]{Computer systems organization~Embedded systems}
\ccsdesc[300]{Computer systems organization~Redundancy}
\ccsdesc{Computer systems organization~Robotics}
\ccsdesc[100]{Networks~Network reliability}

%
%

%
%


\keywords{Medical Information Retrieval, Negation Detection, Natural Language Processing, Negative Feedback}

\section{Introduction}
Clinical decision making, especially in the case of rare diseases, can be a challenging task that involves making the right diagnosis, finding an appropriate test, or suggesting a treatment plan. Finding relevant information for the the wide variety of health problems physicians encounter on a day-to-day basis is difficult and time-consuming. Due to the exponential increase in the amount of annually published research articles, manually identifying the most important and relevant texts becomes infeasible.

State-of-the-art retrieval models applied to clinical decision support settings rely on full-text indices of bio-medical literature and use the textual content of the patient's health record to construct queries. While these models were originally designed with keyword search interaction in mind, medical case narratives are maintained in natural language, resulting in significantly longer queries than what we are used to in Web search settings. As an example, in our case study the average query length after removing stop words was 57.3 words. 

Besides their mere length, negations represent a particularly challenging aspect of natural language queries. Consider the following example, taken from Topic 1 of the TREC 2014 Clinical Decision Support (CDS) track: \textit{``She denies smoking, diabetes, hypercholesterolemia, or a family history of heart disease.''}. The clinical practitioner encodes explicit knowledge of the absence or invalidity of a range of conditions or findings but our term-based retrieval model readily uses the entire negated passage as query terms. This inappropriate use of the carefully curated clinical narrative results in measurable detriments in retrieval performance. We quantify this effect by comparing two sets of TREC 2014 \& 2015 CDS track case reports: Those containing no negations $D_{+}$ (28 reports), and those containing at least some negated information $D_{-}$ (32 reports). We find a clear negative impact of the presence of negated terms on the retrieval results. Both nDCG (21.3\% improvement) and P@10 (10.8\% improvement) were significantly higher for $D_{+}$ than $D_{-}$.

This observation is not just limited to small academic collections such as the TREC corpus, but also holds in real-world clinical environments. Chapman \textit{et al.}~\cite{chapman2001evaluation} find between 39\% and 83\% of all clinical observations to be described in a negated form.

In this paper, we empirically compare state-of-the-art query filtering techniques as well as novel query-adaptive retrieval models, that actively use negated terms as negative relevance feedback. Our investigation is based on corpora and relevance judgements of the TREC 2014 \& 2015 Clinical Decision Support Track and highlights the merit of the proposed methods.

\section{Background}
This study builds on previous findings from both automatic negation detection in natural language processing and negative relevance feedback for retrieval models. The following paragraphs summarize the most closely related developments in both fields.

Rokach \textit{et al.}~\cite{rokach2008negation} provide an extensive overview of negation detection methods for medical narrative reports. The existing body of previous work can be categorized into knowledge engineering and machine learning based approaches.  We will discuss one representative example per category. Chapman \textit{et al.}~\cite{Chapman2001301} propose NegEx, a regular expression based algorithm to detect negated findings in radiology reports. Testing this algorithm on 1235 findings and diseases in 1000 sentences taken from discharge summaries, NegEx achieved a specificity of 94.5\% and a sensitivity of 77.8\% percent. As an example of machine-learned negation detectors, Agrawal \textit{et al.}~\cite{agarwal2010biomedical} present a conditional random field model, designed to detect negation cues and their respective scopes. The model is trained on the publicly available BioScope corpus~\cite{vincze2008bioscope}. This approach outperformed NegEx with F1-scores of 98\% for detecting cues and 95\% for detecting scopes. 

The field of information retrieval has long-standing experience in using feedback of (pseudo) relevance in the retrieval process~\cite{rocchio1971relevance}. However, explicit non-relevance information has been shown to be more difficult to incorporate. Wang \textit{et al.}~\cite{wang2008study} investigate different methods to improve retrieval accuracy for difficult search queries using negative feedback. Their work covers both language and vector-space models, as well as a number of heuristics for negative feedback. In the \textit{Score Combination} strategy, a positive query representation $Q$ and a negative query representation $Q_{neg}$ are maintained separately. The scores for a given document are computed for both query representations and then combined for the final result.

Previous approaches for using negations in medical IR have focused on removing negated terms completely. Averbuch \textit{et al.}~\cite{Auerbuch2004} were able to improve F-Scores by 8.28\% on average by removing negated UMLS-Terms from queries. Even though this approach has shown to improve retrieval results, a lot of information is lost by altogether filtering negated terms from the query. 

Limsopatham \textit{et al.}~\cite{limsopatham2012exploiting} propose NegFlag, a more nuanced approach which handles negations in medical information retrieval by introducing a new term representation and exploiting term dependence. NegEx is used to detect negations while indexing. Terms, which are identified as negated, are replaced by the original term to which the prefix \textit{``n0''} is prepended. By doing so, they avoid incorrectly returning a document, in which a given positive query term appears in a negated context. Additionally, they reduce the relevance score of a document, if it contains the negated form of a  pair of neighboring positive query terms. The focus of this study is set on the negations occurring in the documents. Negations in the query phrase are treated in a boolean model fashion, \textit{i.e.}\ documents containing the negated term are excluded from the results.
Their analysis was based on the 34 topics from the TREC 2011 Medical Record track \cite{voorhees2012overview}. The goal of this track is to retrieve relevant electronic health records, for a given free text search query. This is decidedly different from the TREC CDS task, where the goal is to retrieve relevant biomedical articles for a given health record.
As compared to a baseline system which does not take negative contexts into account, this approach yielded a 2.9\% relative improvement in P@10. 

Koopman \textit{et al.}\ \cite{Koopman2014} examine the impact of negations using the data of the TREC 2011 \& 2012 Medical Records Track. Following~\cite{limsopatham2012exploiting}, they combine the scores for normal and negated content into a final score for a given query and document pair. They empirically find a negative impact of the presence of negations on retrieval results. Contrary to their assumption that negative content should always be penalized in queries, they show that some queries benefit from the inclusion of negated information. Using this method, they achieve a 10\% improvement of P@10 as compared to a baseline, which does not take negations into account, but the improvement over the complete removal approach is insignificant.

\newpage
\section{Case Study}
\subsection{General Setup}
Our empirical investigation is based on the TREC 2014 \& 2015 Clinical Decision Support track document collection. The corpus consists of an open access subset of PubMed Central, an online repository of biomedical literature, as well as 60 artificial, idealized medical case reports, created by experts at the U.S.~National Library of Medicine. As per the track's guidelines, our retrieval experiments use the full text narrative of these reports as queries.

The document collection is indexed using Apache Lucene, with default settings. After the inspection of a broad method and parameter sweep, we rely on an Okapi BM25 retrieval model~\cite{robertson1995okapi} which delivered consistently strong results.

For our queries, we extract the \textit{description} of the provided topics. We apply lower-casing and remove stop words. In the following, we will utilize four different versions of queries:
\begin{enumerate}
    \item The full description \textbf{($Q_{full}$)}
    \item The description, from which all negated sub-sentences were removed \textbf{($Q_{pos}$)}
    \item The negated sub-sentences
    \textbf{($Q_{neg}$)}
    \item The description, in which \textit{``[nx]''} is prepended to words appearing in a negated scope \textbf{($Q_{tagged}$)}
\end{enumerate}

Note that with \textit{``negated terms''} or \textit{``negations''}, we describe the entire negated sub-sentences from here on out. 

As a proof of concept, negations and their scopes were initially annotated manually. Empirical comparison with NegEx~\cite{Chapman2001301} showed only negligible differences that did not have a noticeable effect on retrieval performance.

\subsection{Methods}
\subsubsection{Filtering}
The traditional way of addressing negations in natural language queries, as investigated by~\cite{Auerbuch2004} simply removes negated sub-sentences from the query. The score for a document $D$ and query $Q$ is computed as:

\begin{displaymath} S(Q,D) = S(Q_{pos},D)
\end{displaymath}

where $S(Q,D)$ is the BM25 score of document $D$ for query $Q$.

\subsubsection{Score Combination}
While the filtering approach to negation handling has been shown to perform well in practice, intuition mandates that making explicit use of the information contained in the negation should be beneficial. Inspired by~\cite{Koopman2014}, we rely on~\cite{wang2008study}'s score combination method that computes the relevance score for query $Q$ and document $D$ as:
\begin{displaymath} S_{combined}(Q,D) = S(Q_{full},D) - \beta \times S(Q_{neg},D)
\end{displaymath}
 We adapt this method to our needs by constructing $Q_{neg}$ from the negated query terms, instead of using negative document examples. We denote the number of terms in the current query as $n_{full}$. To avoid assigning too much weight to negative terms, if they occur infrequently, we set $\beta$ in the following, empirically determined manner by optimizing for P@10:

\begin{displaymath}
\beta = -0.0001638 \times n_{full}^2 + 0.04631 \times n_{full} - 1.207.
\end{displaymath}

\subsubsection{Negation Tagging}

The key idea of our novel Negation Tagging method is to match the contexts of query terms and the contexts of the same terms in documents. This is based on the assumption that a document might be relevant for a negated query term, if it also contains the term in a negated form.
Thus, following~\cite{limsopatham2012exploiting}, negated terms and their scopes are tagged in the same way in both the queries and the documents. Specifically, we add the prefix \textit{``[nx]''} to negated terms. Instead of demoting documents containing opposite contexts in the spirit of Limsopatham \textit{et al.}, we increase their score if they contain positive forms of negated query terms. A document can thus still be deemed relevant, even if it contains the positive form of a negative query term. This is achieved by expanding the query by the untagged form of each originally negated query term. If a given query contains no negations, the baseline algorithm is applied. Otherwise, the relevance score for a given query $Q$ and document $D$ is computed as: 

\begin{displaymath} S(Q,D) = S(Q_{tagged},D_{tagged}) + \beta \times S(Q_{neg},D_{tagged})
\end{displaymath}
where $D_{tagged}$ is the tagged document. To avoid assigning too much importance to these expansion terms, their term weights are set to $\beta = 0.3$, a setting that was empirically determined to perform well.

\subsection{Results}
As the number and extent of negated phrases among the provided queries is relatively low (on average 3.75 words per 57-term query), the maximal potential impact of our methods is limited. Nevertheless, score combination and negation tagging not only outperform the baseline in P@10, but also improve on the established negation filtering strategy. (see Table~\ref{table:results}). Importantly, this improvement is achieved without negatively impacting the performance of queries without negations.

While negation tagging and score combination yield comparable overall improvements, when considering the query with the most significant amount of negated information (Topic 1 from the 2014 CDS track), score combination achieves the most pronounced gain of up to 300\% relative improvement (see Table~\ref{comparison_topic1}).

In line with~\cite{Koopman2014}'s findings, we also note that a few topics benefit from the boosting of negated terms when applying score combination. We could, however, not find a systematic connection between those queries for which the optimal $\beta$ takes a negative value. These results further emphasize the strength of negation tagging, which works well with a non-adaptive $\beta$.

As mentioned above, we also split up our topics into two sets of queries, those containing no negations $D_+$ (28 reports) and those containing at least some negated information $D_-$ (32 reports). We find a clear negative impact of the presence
of negated terms on the retrieval results. All of the considered metrics were significantly higher for the group without negations: infAP (23.4\% better), nDCG (21.3\% better), P@10(10.8\% better) and RPrec (10.7\% better). To measure how close the methods under consideration get to closing this performance gap, we apply the algorithms to $D_-$ separately and then compare their performance to the baseline results for $D_+$. The results of this examination are displayed in Table~\ref{table:groups}. Remarkably, Negation Tagging and Score Combination reduce the difference in P@10 by 65.4\%. Namely, these methods improve P@10 by 7.1\% as compared to the baseline for $D_-$. The retrieval performance as measured by the other metrics is maintained. 

\begin{table}[]
\centering
\caption{Comparison of all methods on all topics.}
\label{table:results}
\begin{tabular}{|l|l|l|l|l|}
\hline
                   & P@10    & NDCG    & infAP   & RPrec   \\ \hline
Baseline           & 0.325   & 0.3072 & 0.0978  & 0.1597  \\ \hline
Negation Filtering & 0.3283  & 0.3058 & 0.0966 & 0.1598 \\ \hline
Score Combination  & 0.3367 & 0.3063  & 0.0972  & 0.1581 \\ \hline
Negation Tagging   & 0.3367  & 0.3054 & 0.0974  & 0.1581  \\ \hline
\end{tabular}
\end{table}

\begin{table}[]
\centering
\caption{Between group comparison of methods. The rows are labeled with the algorithm applied and the set of topics under consideration.}
\label{table:groups}
\begin{tabular}{|l|l|l|l|l|}
\hline
             & P@10       & NDCG       & infAP      & RPrec      \\ \hline
Baseline D+ & 0.3429    & 0.3389     & 0.1088     & 0.1684    \\ \hline
Baseline D-  & 0.3094    & 0.2794     & 0.0882     & 0.1521     \\ \hline
Negation Filtering D-        & 0.3188 & 0.2766 & 0.0856 & 0.1526   \\ \hline
Score Combination D-        & 0.3313     & 0.2778   & 0.087  & 0.1491  \\ \hline
Negation Tagging D-        & 0.3313 & 0.2761 & 0.0873   & 0.1491 \\ \hline
\end{tabular}
\end{table}

\begin{table}[]
\centering
\caption{Comparison of Methods, Topic 1. Best results marked in bold.}
\label{comparison_topic1}
\begin{tabular}{|l|l|l|l|l|}
\hline
                    & P@10 & NDCG   & infAP  & RPrec  \\ \hline
Baseline            & 0.1  & 0.2664 & 0.0382 & 0.1341 \\ \hline
Negation Filtering  & 0.3  & 0.2252 & 0.0359 & 0.1341 \\ \hline
Score Combination   & \textbf{0.4}  & \textbf{0.2805} & \textbf{0.0499} & \textbf{0.1341} \\ \hline
Negation Tagging    & 0.2  & 0.2480 & 0.0363 & 0.1220 \\ \hline
\end{tabular}
\end{table}

\subsection{Limitations}

Clearly, the expressiveness of the results presented here is limited due to the small sample size as well as the relative brevity of case reports. Real-world medical case narratives often span multiple pages or volumes as the patient history unfolds across years of treatment. For instance, the topics for the current TREC 2016 CDS track belong to authentic critical care patients and contain on average 94.8 words out of which 6.13 are negated. In comparison,  our artificial case reports contain 57.3 words on average (3.75 of them negated).

\section{Conclusion}

Making use of negative information is critical for retrieving documents in clinical contexts. In this paper, we have laid out how automatic negation detection output can be utilized by actively discounting documents containing negated query terms or introducing new term representations in both queries and documents. Our case study indicates that these approaches are more promising than ad-hoc removal of negated terms. Empirical results show that our negation tagging methods are able to eliminate the negative impact of negations on early precision almost completely, while maintaining the retrieval quality for other metrics. The results for negation-heavy queries furthermore indicate that a score combination approach, with an appropriate weighting parameter, might yield even better results.

There are several interesting research questions that we aim to address in the future: (1) This work studied a small academic sample of carefully curated artificial case reports. In the future, it will be mandatory to investigate the generalizability of our findings to real-world collections of considerable size. (2) Similarly, we aim to investigate the effect of going beyond the currently studied short and artificial patient records towards longer clinical narratives. (3) Finally, in the future, adaptive choices of $\beta$ should account for the actual importance of negated terms and not just their relative length. An adaptive way of determining the sign and magnitude of $\beta$ has to be found.

%
\bibliographystyle{abbrv}
\bibliography{sigproc}  
%
%

\end{document}